\documentclass{iaus}
\usepackage{graphicx}
    
\newcommand{\logt}{log T$_{eff}$}
\newcommand{\logl}{log L/L$_\odot$}

\begin{document}

\title[IAU Symposium 258.~~Young Star Ages] 
{Age-Related Observations of Low Mass Pre-Main and Young Main Sequence
Stars (Invited Review)}

\author[Lynne A. Hillenbrand]{Lynne A. Hillenbrand$^1$}

\affiliation{$^1$California Institute of Technology \\ MC 105-24,
Pasadena, CA 91125 (USA) \\ email: {\tt lah@astro.caltech.edu}}

\pubyear{2009}
\volume{258} 
\pagerange{1--8}
\setcounter{page}{1}
\jname{The Ages of Stars}
\editors{E.E. Mamajek, D.R. Soderblom, R.F.G. Wyse, eds.}

\maketitle

\begin{abstract} This overview summarizes the age dating methods
available for young sub-solar mass stars.  Pre-main sequence age
diagnostics include the Hertzsprung-Russell (HR) diagram,
spectroscopic surface gravity indicators, and lithium depletion;
asteroseismology is also showing recent promise.  Near and beyond the
zero-age main sequence, rotation period or vsin$i$ and activity
(coronal and chromospheric) diagnostics along with lithium depletion
serve as age proxies. Other authors in this volume present more detail
in each of the aforementioned areas.  Herein, I focus on pre-main
sequence HR diagrams and address the questions: Do empirical young
cluster isochrones match theoretical isochrones?  Do isochrones
predict stellar ages consistent with those derived via other
independent techniques? Do the observed apparent luminosity spreads at
constant effective temperature correspond to true age spreads?  While
definitive answers to these questions are not provided, some methods
of progression are outlined.
\end{abstract}

\firstsection 
\section{Techniques for Assessing Young Star Ages}

Standard stellar age dating techniques can be divided in to those
which are purely empirical in nature and those which are more
theoretically grounded. The former include measurements related to
stellar kinematics and cluster membership, stellar rotation as derived
from periodic photometric modulation or spectroscopic absorption line
broadening, stellar chromospheric activity as measured by
e.g. fractional Ca II H\&K or H$\alpha$ line luminosity, stellar
coronal activity as measured by soft x-rays, lithium depletion trends,
and age-metallicity as well as age-velocity dispersion relations.
Most of the empirical correlations have dependencies on the stellar
mass in addition to the age effect which is sought, adding necessary
complication to any analysis.  The latter methods, those referenced
more directly to theory, include location in the Hertzsprung-Russell
(HR) diagram relative to calculated isochrones, theoretical nuclear
burning as traced though e.g.  lithium abundances, and
Asteroseismological constraints.

Each of the above diagnostics is in principle also applicable in the
pre-main sequence phase of stellar evolution, as well as to the
zero-age and young main sequence phases.  However, many of them are
significantly diminished in value as quantitative pre-main sequence
age indicators due to ``saturation" or ``degeneracy" effects.
Specifically, the coronal and chromospheric activity indicators which
generally decline in strength with advancing main sequence age
(e.g. Mamajek \& Hillenbrand, 2008) are either very shallowly
dependent on, or constant with, age for solar-type pre-main sequence
stars.  Rotational behavior likewise deviates from the monotonic
spindown characteristic of main sequence angular momentum evolution,
with significantly higher dispersion in rotation properties observed
at ages younger than $\sim$200 Myr. This is explained as remnant
behavior related to earlier interaction of the stellar magnetosphere
with the primordial circumstellar disk, specifically star-to-star
variation in the time scale for star-disk coupling. Kinematics and
cluster membership still apply as age diagnostics in the pre-main
sequence phase, and are employed in the same ``guilt by association"
manner as utilized for main sequence clusters. However, the absolute
age dating is more difficult.

Another method that is sometimes used as a relative age dating
technique for the very youngest stars is the fractional infrared
excess luminosity, or the shape of the mid-infrared spectral energy
distribution.  While it is true that the vast majority of stars with
remaining detectable primordial circumstellar dust are younger than
$\sim$10 Myr, and that the vast majority of stars with so much dust
that they are seen via scattered light or are still partially or
totally self-embedded are younger than $\sim$1-2 Myr, there is no
evidence for a monotonic relationship between circumstellar dust
characteristics and absolute stellar age.  On the contrary, there are
strong arguments for significant {\it dispersion} in the amount of
circumstellar dust (and gas) among stars aged less than $\sim$10 Myr
-- even those located in the same cluster or association.  Thus, while
stellar youth is certainly indicated by the presence of circumstellar
material, the quantitative use of circumstellar properties as stellar
chronometers is not recommended and will not be discussed further
here.

We are thus left with three stellar age dating methods that are both
applicable and increasingly well-calibrated at young
-- pre-main sequence and zero-age main sequence --  ages:  
(1) the theoretical HR diagram or extinction-correction color-magnitude
diagram, 
(2) inferences of log $g$ vs. \logt\ from spectra, and 
(3) lithium abundance measurements and depletion trends.
There is also some promise from (4) asteroseismology but this method 
has not yet proved itself.  In what follows I discuss each of these four
techniques and recent results. 

\section{Stellar Age Dating in Regions of Recent Star Formation}

Stars form within giant molecular clouds that become unstable to
fragmentation and subsequent gravitational collapse to produce:
stellar clusters, multiple star systems, and individual stars.  The
time for an individual protostar to collapse is related to the local
sound speed, and is expected to be 0.1-0.2 Myr (Shu, et al. 1987).  On
larger spatial scales, two main theories of star formation suggest
different regulating phenomena and therefore time scales for the
start-to-finish process of star formation in a molecular cloud.
Regulation by quasi-static ambipolar diffusion processes takes
$\sim$3-10 Myr (Shu, 1977, Mouschovias, 1976) while regulation by
turbulence dissipation occurs on the dynamical time scale of only
$\sim$0.5-few Myr (Ballesteros-Paredes et al. 1999, Elmegreen 2000).

We can hope to probe the relative importance of these physical
processes by studying the mean ages and detailed age distributions in
regions of current and recent star formation.  Our main tools are
those mentioned above: stellar bolometric luminosities and the HR
diagram, spectral diagnostics of stellar surface gravity, and
measurements of Li I 6707 \AA\ abundances.  The bulk of my discussion
concerns HR diagrams.

\subsection{HR Diagrams}

A good case study that informs our understanding of stellar ages and
age spreads in star forming regions is the Orion Nebula Cluster (ONC).
Hillenbrand (1997) published a synthesis of existing and new
photometry and spectroscopy in this region, enabling the location of
over 900 stars on the theoretical HR diagram.  Now, new and better
photometry from HST/ACS and ESO/WFI as well as over 600 new optical
spectral types from WIYN/Hydra, Palomar/Norris, and Keck/LRIS are
available.  Also, recent estimates of the ONC distance place the
cluster $\sim$15\% closer than previously accepted values.  Further,
we now have a better understanding of the photometric variability
trends and amplitude ranges of individual ONC stars, enabling use of
median photometry.  The improved data along with revisions in our
understanding of the temperature and bolometric correction scales
appropriate for young pre-main sequence stars makes it worth
revisiting the finding of a substantial luminosity spread in the
Hillenbrand (1997) study.

First attempts at revision are presented by Da Rio et al. (2009),
considering very carefully the subtleties of young star de-reddening
and the effects of accretion, and Reggiani et al. (poster at this
meeting), considering only the least photometrically variable stars.
These authors find essentially no reduction in the $\sim$1.5 dex
luminosity spread [or $\sigma$(\logl) $\approx 0.55$ dex at fixed
\logt] characteristic of the lower quality and single epoch
Hillenbrand (1997) data.  Thus, observational errors and biases, and
known causes of scatter do not appear to be the main culprit in
creating the large luminosity spreads that are observed in the ONC HR
diagram.

Indeed, such apparent luminosity spreads have been seen for decades in
young cluster and association HR diagrams.  Literature from the 1960s
and 1970s, e.g. the venerable Iben \& Talbot (1966) and Ezer \&
Cameron (1967) showed them.  Such early comparisons between data and
theoretical pre-main sequence isochrones are primarily responsible for
long standing paradigms such as ``molecular clouds form stars for
about 10 Myr" and ``circumstellar disks last about 10 Myr."  Although
the above statements have been modified with better data and modern
interpretation, the evidence for cluster luminosity spreads has
persisted.  Thus the questions remain: Are the apparent luminosity
spreads real?  Do they indicate true age spreads?  Can we use them to
infer star formation histories?

\subsubsection{HR Diagram Methodology}

Before embarking on these questions of HR diagram interpretation, it
is important to review how stars are located in the HR diagram based
on observational data and available techniques, as well as the
accompanying complications to such procedures.

In practice, a spectral type determined at blue optical (BV), red
optical (RI), or near-infrared (YJHK) wavelengths is used along with a
spectral-type- to-effective-temperature conversion to set the abscissa
in the HR diagram.  Photometry within some subset of optical or
near-infrared bands is used along with the spectral type to calculate
and correct for reddening, and then a bolometric correction
appropriate to the spectral type is adapted in order to calculate the
ordinate of bolometric luminosity.

Complications to this standard process that are unique to young stars
include effects related to the ubiquitous presence of circumstellar
disks for some portion of the early pre-main sequence (see Meyer, this
volume).  Accretion from the disk on to the star creates a hot excess
which makes blue photometry ``too blue," while thermal plus accretion
emission from the inner disk makes red photometry ``too red."  Both
phenomena confuse de-reddening procedures.  The potential existence of
both blue and red excess means that, in fact, there may be no truly
photospheric wave band at which to apply bolometric corrections to the
reddening corrected photometry.  Furthermore, some young sources are
not seen directly, but via light scattered through circumstellar disks
or envelopes which leads to significant luminosity underestimation.
For example, all Taurus scattered light sources sit on or below the
zero-age main sequence. The extent to which scattered light affects
other systems, in which it is not known from spatially resolved
images, is unknown.  Luminosity effects resulting from typical
parameter distributions star plus disk systems were modeled by Kenyon
\& Hartmann (1990) who found induced luminosity deviations relative to
non-disked stars of $\sigma$(\logl) $<$ 0.2 dex.  Other concerns for
HR diagram construction at young ages include generally large values
of visual extinction ($>$1-10 mag), uncertainty regarding the
appropriate extinction law, and significant photometric variability at
typically $<$0.2 mag levels though $>$1 mag in more extreme cases.

Additionally present for the young stars are the usual complications
affecting all HR diagram determinations.  These include random errors
due to spectral type and photometric uncertainties, and systematic
errors deriving from unresolved multiplicity that result in luminosity
overestimates (e.g. Simon et al. 1993).

\subsubsection{Checks on Methodology}

Checks on our ability to locate young stars in the HR diagram are
provided by binary and higher order multiple systems, whose components
we expect to be coeval.  Previous work in this area includes that by
Hartigan et al. (1994), White et al. (1999) Prato et al. (2003), and
Ammler et al. (2005).  Recently, Kraus \& Hillenbrand (2009) have used
more modern temperature and luminosity pairs based on improved
photometry and spectral types from high spatial resolution data to
determine that, indeed, the binaries and higher order multiples in the
Taurus-Auriga region are more coeval than random pairings of member
stars.  However, while some multiple systems lie on theoretical
isochrones (within the errors), others are significantly mismatched.
It is unclear at present whether the observed effects can be
attributed to random or systematic errors, or if they indicate true
non-coevality, but the result should be kept in mind as we proceed to
discuss luminosity spreads in clusters as a whole.

Totally independent checks on \logl\ and \logt\ conversions via the
theoretical HR diagram to stellar {\it masses} come from comparison of
such predictions to dynamical mass measurements.  However, similarly
independent checks of \logl, \logt\ to stellar {\it age} predictions
are more difficult to develop.  At best, we can demand that the
theoretical isochrones are parallel to observed cluster sequences,
similar to the expectations for binaries and higher order multiples.
We can also hope for consistency with other techniques, such as e.g.
turn-off ages for the higher mass stars in the same cluster, surface
gravity measurements, lithium abundance determinations, etc.

\subsubsection{HR Diagram Theory}

As we aim to use HR diagrams to gain knowledge about absolute stellar
ages, in addition to investigating the evidence for or against spreads
in age, it is important to discuss in brief the calibrating theory for
the stellar age determinations.

As detailed in Hillenbrand et al. (2008), there are significant
systematic effects between available theoretical predictions of
pre-main sequence evolution.  Specifically, the trend at sub-solar
masses is that for a given location in the HR diagram, the youngest
ages are those inferred from the D'Antona \& Mazzitelli (1994, 1997
and 1998 update) theory with increasingly older ages predicted by Yi
et al. (2001, 2003, 2004), Swenson et al. (1993), Palla \& Stahler
(1993, 1999), Siess et al. (2000), and finally the Baraffe et
al. (1995, 1998) theory predicting the oldest ages.  Age differences
between these various track sets for the same \logl, \logt\ pair rise
to $\sim$0.75 dex at the youngest ages!

Furthermore, with all isochrone sets, trends of stellar age with
stellar mass are present, as widely reported in observational papers.
Along any empirical isochrone $<$50-80 Myr, the higher mass stars
generally appear older than the lower mass stars.  This suggests
either that we are still missing physics associated with the initial
appearance of stars in the HR diagram, or that the mass-dependent
physics of stellar interiors is still not adequately understood.  As
detailed by Palla (this volume), additional complications to pre-main
sequence evolutionary theory such as the effects of initial conditions
(i.e. the ``birthline"), disk accretion, stellar magnetic fields, and
stellar rotation may explain some of the observed dispersion between
observationally derived and theoretically predicted effective
temperatures and luminosities.

\subsubsection{Empirical Results for Young Clusters}

\begin{figure}[t]
\begin{center}
 \includegraphics[angle=-90, width=4in]{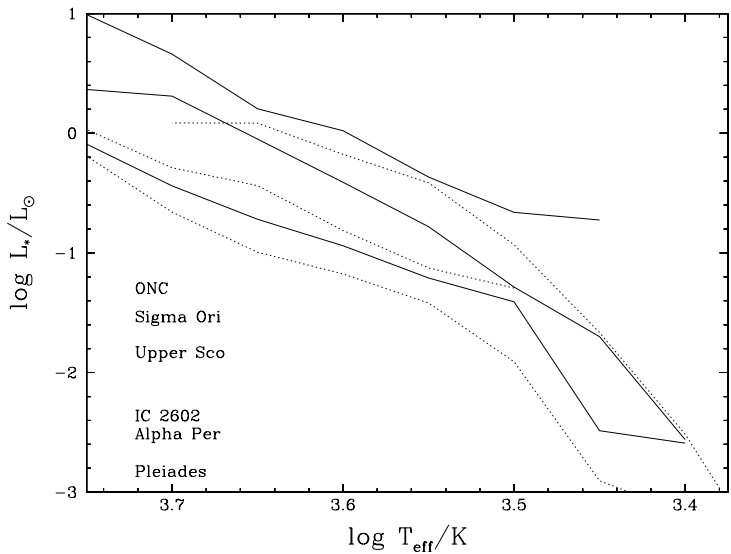} 
 \caption{A sequence of empirical isochrones for representative young clusters 
 and associations in the solar neighborhood.  
 Shown is the median stellar bolometric luminosity as a function of 
 stellar effective temperature for member star and brown dwarfs. 
 Ordering in the legend corresponds to relative luminosity at \logt=3.6.}
   \label{fig:f1}
\end{center}
\end{figure}

I apply the methods outlined above and return now to the question of
how well pre-main sequence clusters compare to theoretical isochrones.
Shown in Hillenbrand et al. (2008) are the \logl\ and \logt\
distributions -- represented as the mean and 1-$\sigma$ luminosity
vs. binned effective temperature -- for over 25 young clusters,
associations, and currently active star forming regions within 500 pc
of the Sun.  I take the mean luminosity with effective temperature
sequence in each region as an empirical isochrone.  As illustrated
comparatively in Figure~\ref{fig:f1}, there is good representation
among this sample of stellar populations having ages from $<1$ Myr to
just over 100 Myr.  The median luminosities at fixed effective
temperature span approximately 1.5 dex.

I quantify the luminosity {\it spreads} in each region by calculating
the distribution of $\Delta$\logl = [log $L_{observed}(i)/L_\odot -$
log $L_{median} ($log $T_{eff})/L_\odot]$, or the deviation of
individual luminosities from the median value appropriate for the
temperature.  These distributions, also, are illustrated in
Hillenbrand et al. (2008).  In most cases, Gaussian fits appear to
describe adequately the luminosity deviations, suggesting that random
processes are the dominant contributor to the luminosity spreads.

In detail, the fitted dispersions to the $\Delta$\logl\ distributions
for somewhat older ($>$30 Myr) near-main sequence young clusters, such
as the Pleiades, $\alpha$ Per, IC 2602, IC 2391, and the Tucanae /
Horologium Association, are low with $\sigma(\Delta$\logl) = 0.10-0.15
dex.  We can take this as the typical luminosity dispersion that may
be expected from the HR diagram placement methods described above.
Towards younger ages, however, empirical dispersion increases
substantially with $\sigma(\Delta$\logl) = 0.2-0.6 dex for clusters
younger than 3-10 Myr.  These spreads may be compared to the 0.15-0.25
dex dispersions estimated as plausible by Hartmann (2001) for young
stellar populations in which significant accretion luminosity is
present, and the even smaller spreads discussed by Burningham et
al. (2005) as characteristic of young variables. For pre-main sequence
contraction going roughly as L $\propto \tau^{-2/3}$, the {\it
implied} age dispersion from literal interpretation of observed
luminosity dispersion is then $\sigma$(log $\tau) \propto
1.5~\sigma($log $L)$.

However, it is not only the Gaussian width that is important to assess
in considering cluster luminosity dispersion. Rather, it may be the
subtle deviations from pure Gaussianity that convey the important
information apropos, e.g. star formation history of a region, or other
factors such as binary properties of the sample.  Monte Carlo
simulation of the luminosity distributions that accounts for these
various details can help illuminate the important effects, and is
discussed in a later section.

At this point I would like to (re-)emphasize that before any apparent
luminosity dispersion is considered real, that observational fidelity
must be verified so as to minimize any contaminating effects to the
already complex interpretation of the luminosity spread phenomenon.
First, we should ensure that we are considering only certain cluster
or association members and regions that are not confused by superposed
episodes of star formation.  Next, we should strive to obtain
exquisite photometry and high quality spectroscopy so as to reduce the
influence of random observational errors.  We should account for
possible scattered light (causing luminosity underestimates) in young
regions and multiplicity (causing luminosity overestimates) in all
regions; although these both are systematic effects, they apply to
only some portion of the population and therefore contribute to
apparent luminosity spreads.

In summary, only pristine samples and the best data should be used in
probing luminosity distributions.  I turn now to discussion of
potential correlates with $\Delta$\logl.

\subsubsection{Independent Observational Checks}

We can test the reality of the observed apparent luminosity spreads
via their confirmation by independent observational means.
Specifically, we can look for correlations between $\Delta$\logl\ and
surface gravity indicators or lithium abundance trends.  Further, we
can take advantage of the asteroseismological checks that have
recently come to fruition for pre-main sequence stars in certain mass
regimes.  In addition to the discussion below, I refer the reader to
clever techniques pioneered by Jeffries (this volume) and Naylor (this
volume) which also provide checks on the observed apparent luminosity
spreads in young clusters.

\subsection{Surface Gravity Diagnostics}

Low mass stars have a number of surface gravity sensitive spectral
features in the red optical wavelength range -- that most often used
to classify such objects in modern studies -- with others available at
near-infrared wavelengths but not discussed here.  For stars with
spectral types later than $\sim$M2, the CaH 6975~\AA\ band and the Na
I 8183,8195~\AA\ doublet lines are surface gravity sensitive at ages
younger than $\sim$20-30 Myr (Schiavon et al. 1995, Slesnick et
al. 2006).  Towards later spectral types, beyond $\sim$M6 and
extending well into the L types, the K I 7665,7699 doublet lines, and
several VO bands are additionally useful surface gravity diagnostics
at ages younger than $\sim$ 100 Myr (Steele \& Jameson 1995,
Kirkpatrick et al. 2008).

\begin{figure}[t]
 \vspace*{-5.0 cm}
\begin{center}
\includegraphics[angle=-90, width=\textwidth]{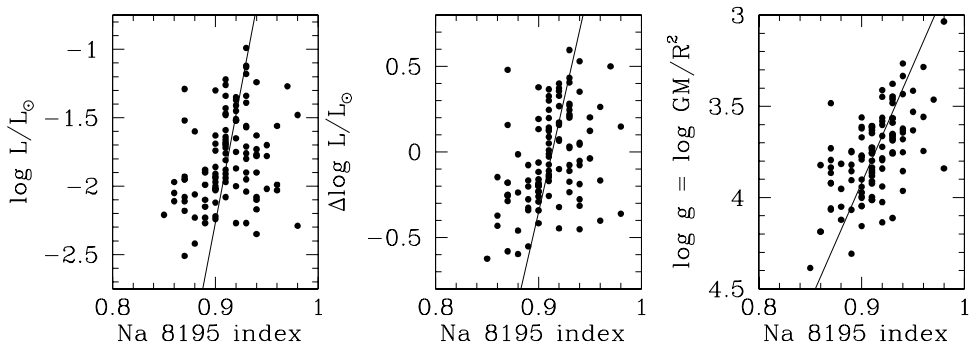} 
 \caption{
 Based on data presented in 
 Slesnick et al. (2008) for M4-M7 stars in the Upper Sco
 region. The left, middle, and right panels correlate
 \logl\ (stellar bolometric luminosity),
 $\Delta$\logl\ (deviation from mean luminosity normalized
 to effective temperature), and
 log $GM/R^2$ (surface gravity) computed from the pre-main sequence
 \logl\ and \logt\ location in the HR diagram -- all 
 with the surface gravity sensitive Na I 8190~\AA\ spectroscopic
 index defined by Slesnick et al.  
 Correlation coefficients and the linear least squares fits are poor 
 for the left and middle panels, but -0.6 (inversely correlated) 
 in the right panel with 0.27 dex rms for the displayed fit of 
 [log $g = (-12.9 \pm 0.7) \times Na I + (15.5 \pm 0.6)$]. 
 }
   \label{fig:f2}
\end{center}
\end{figure}

One test of the reality of the observed apparent luminosity spreads in
young clusters is whether there is any correlation between
$\Delta$\logl, the deviation from the median cluster luminosity
normalized for effective temperature, and a quantitative index of
surface gravity.  With $L \propto R^2T^4_{eff}$ and $g \propto M/R^2$,
we expect for low mass stars of constant mass contracting along
Hayashi (roughly constant temperature) tracks that
\logl\ and log $g$ should be inversely correlated. Slesnick et al. (2008) 
demonstrate for stars in the Upper Sco region that objects with the same or 
similar measures of the surface gravity sensitive Na I 8190~\AA\ spectral index 
can have a broad range of luminosity.  Although the median age 
of the cluster is $\sim$5 Myr, individual stars with the same Na I
index have ages predicted based on their luminosities from $<$3 to $>$14 Myr. 
If these luminosity-based ages are to be believed, we would expect 
corresponding differences in the Na I index.  

What is observed is shown in Figure~\ref{fig:f2}. Although there is
significant scatter, it does appear that the surface gravities implied
from HR diagram location do correlate in the expected way with a
completely independent (spectroscopic) indicator of surface gravity.
However, neither the straight \logl\ nor the $\Delta$\logl\ values
exhibit similar correlation.  Luminosity based ages for individual
stars thus still warrant considerable skepticism, and cautions remain
against uncritical assessment of observed apparent luminosity
dispersion as true age dispersion.  However, the quantitative results
shown here do imply non-zero spread in both luminosity-based surface
gravity and spectroscopic surface gravity, and thus by implication
perhaps age at the several Myr level.

\subsection{Lithium 6707 \AA\ Measurements}

Low mass stars burn both deuterium and lithium during the pre-main
sequence evolutionary stages, essentially early steps in the hydrogen
burning set of reactions that take place later on the main sequence.
Contracting objects between $\sim 1-2.5 M_\odot$ undergo lithium
burning processes for only a few tens of Myr to $<$1 Myr, while those
below $\sim 1 M_\odot$ and down to the hydrogen burning limit deplete
their lithium essentially forever, and brown dwarfs burn only
deuterium but never lithium (e.g. D'Antona \& Mazzitelli, 1994; Nelson
et al. 1993).  Lithium depletion trends in young pre-main sequence and
main sequence populations have been used at sub-solar masses to
estimate stellar ages, as discussed elsewhere in this volume.  There
is considerable scatter at constant age (e.g. within clusters) in both
the observed equivalent widths and the derived abundances at constant
mass or spectral type for stars younger than a few hundred
Myr. Physically, this dispersion in surface abundance is likely
related to the dispersion in rotation speeds over the same age range.

For any given star, lithium depletion is monotonically related to
stellar age in the sense that lithium is never created via nuclear
reactions, only destroyed.  A test, therefore, of the reality of
observed apparent luminosity spreads in young clusters is whether
there is any correlation between $\Delta$\logl, the deviation from the
median cluster luminosity normalized for effective temperature, and
lithium abundance.  Palla et al. (2005, 2007) have argued in the case
of a small sample in the ONC that this is indeed the case, with
isochronal and lithium depletion ages agreeing to within 5\% in most
cases based on the models of Siess et al. (2000).  The agreement is
particularly noted for those objects which sit low in the HR diagram
relative to the main locus.  However, it is just these stars which are
suspected of being slightly foreground interlopers, part of the Orion
Ic association which is indeed older and envelops the Orion Id (ONC)
region.  Thus, although intriguing and certainly an excellent way to
test the conundrum of large luminosity spreads, in the particular case
of the ONC there may be other complications which overshadow the main
effect of this comparison.

An interesting case is that of St 34 in Taurus (White \& Hillenbrand,
2005), a near-equal mass binary with both components sitting low in
the HR diagram relative to other Taurus members (isochronal age
$\sim$8 Myr) and also near-fully lithium depleted (depletion age $>$25
Myr).  Otherwise, the star has all the characteristics typical of
classical T Tauri stars: strong H$\alpha$, He I, other
accretion/outflow spectroscopic diagnostics, infrared excess, etc.  It
is thus either an unusually long-lived accretion disk system, or has
had a somewhat unusual radial contraction and very unusual lithium
depletion history.  Another mysterious young object with apparent
lithium depletion age much older than its isochronal age is Beta Pic
group member HIP 112312 A (Song et al. 2002).

The existence of a few potentially anomalous objects like the examples
above not withstanding, the correlation between lithium depletion and
\logl\ should be investigated more broadly.  Recent studies of lithium
in Taurus by Sestito et al. (2008) and in older nearby associations by
Mentuch et al. (2008) provide some of the needed data.

\subsection{Asteroseismology}

Pulsational behavior in the Sun and other stars has offered important
checks on our stellar interior models.  Typically, e.g. in the case of
the Sun as well as near the classical instability strip on the HR
diagram, the pulsation mode is driven by opacity sources (the $\kappa$
mechanism).  Some pre-main sequence stars of intermediate mass lie
near this strip (Marconi \& Palla, 1998) and are being monitored for
pulsations with a good number of detections to date (e.g. Zwintz et
al. 2008 and references therein).  Additionally, there is a prediction
(Palla \& Baraffe, 2005) at the lowest masses that stellar/sub-stellar
interior instability can be driven by deuterium burning ($\epsilon$
mechanism) and also result in observable pulsational behavior.  A {\it
narrow} instability strip that is nearly parallel to the isochrones
offers for these very low mass stars and brown dwarfs strong age
constraints that are totally independent of the HR diagram
-- if pulsators can be found.

Many of the known brown dwarfs in star forming regions such as
Chamaeleon I/II, Lupus, Ophiuchus, Upper Scorpius, IC 348, Sigma Ori,
and the ONC lie near the predicted instability strip.  Candidate
objects are being monitored in thesis work by A.M. Cody (poster
presented at this meeting) for photometric variability to determine
which might be pulsators.  Thus far, interesting variability at the
right amplitudes ($<$0.02 mag) and on the right time scales (several
hours) has been detected through periodogram analysis.

Significant work of a very detailed nature still needs to be conducted
on both the observational side and the theoretical side of pre-main
sequence pulsations.  In principle, however, asteroseismology is a
powerful technique for assessing independently the stellar ages
inferred from HR diagrams.

\section{Simple HR Diagram Simulations}

\begin{figure}[t]
\begin{center}
 \includegraphics[angle=-90, width=\textwidth]{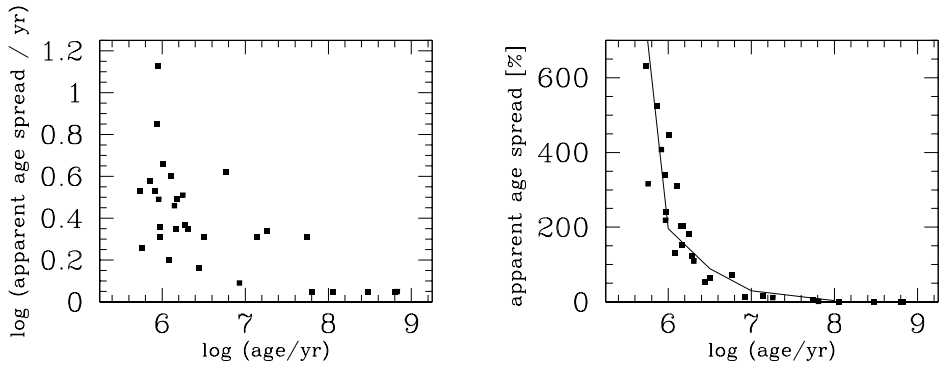} 
\vspace*{-5.0 cm}
 \caption{ Left panel: 1-sigma log age dispersion versus mean log age
 as predicted from the \logl\ and \logt\ data by D'Antona \&
 Mazzitelli (1997/1998).  Other tracks generally yield older ages and
 even larger age dispersions.  Right panel: corresponding percentage
 (linear) age error.  The line indicates the expected age error
 introduced by an imposed luminosity scatter of 0.2 dex, and
 reproduces the observed ``age scatter" for the youngest clusters
 though overpredicts the ``age scatter" for older clusters where the
 actual luminosity spread is indeed lower.  We can conclude that the
 observed luminosity dispersion is comparable to the luminosity spread
 expected from the random and systematic errors suffered during HR
 diagram placement, rather than being dominated by true age spread.  }
   \label{fig:f3}
\end{center}
\end{figure}

Having discussed young cluster HR diagrams and several independent
checks on the observed apparent luminosity spreads, we turn briefly
now to simulation of those luminosity spreads.

We consider as a first simplistic look, a situation in which the
uncertainty ($\sigma$) in individual values of \logl\ which
empirically characterizes the luminosity dispersion in older (main
sequence) clusters, $\sim$0.1-0.2 dex.  We can then propagate such
$\sigma$(log L) values into $\sigma$(log $\tau$) values, and compare
to empirically inferred (from the observed $\Delta$\logl\
distribution, or the apparent luminosity spread) values of $\Delta$log
$\tau$, or apparent age spread.  I show in Figure~\ref{fig:f3} the
predicted trend of $\sigma$(log $\tau$) vs log $\tau$ compared to the
trend actually inferred from young cluster HR diagrams.  As can be
seen, the assumption of luminosity errors typical of those on the main
sequence leads to the expectation that age errors should rise towards
younger ages to $>$100-500\% at ages $<$1-3 Myr.

Slesnick et al. (2008) performed a more realistic Monte Carlo
simulation that projected on to the HR diagram the combined effects of
various errors appropriate for late type pre-main sequence stars.  An
underlying 5 Myr coeval population was masked by: photometric error of
0.025 mag in each observation band (typical of observations), spectral
type error of 1/2 spectral subclasses, (typical of M-types), distance
spread in the population (rendering the model cluster as deep as it is
wide), and stellar multiplicity with 1/3 of the systems being equal
mass binaries (consistent with the mass ratios typically observed for
low mass stars and brown dwarfs).  The simulation resulted in a mean
age and age dispersion totally consistent both visually and
statistically with that found for the low mass population in the Upper
Scorpius region.  In other words, despite the apparent luminosity
spread, no true age spread was needed in order to model the data.

Even more sophisticated Monte Carlo simulations were performed by
A. Bauermeister in undergraduate thesis work. 
The models consider different possibilities for both evolutionary tracks
and star formation histories, along with realistic input error 
and binary distributions in order to simulate cluster age spreads.
The simulations can be analyzed in the same way as empirical data,
e.g. calculating for the resulting HR diagrams
the median luminosity as a function of effective temperature,
the dispersion and the detailed shape of the $\Delta$\logl\ distribution, and
the slope of \logl\ vs \logt\ fit over a limited spectral type range.

Results thus far (as reported previously in Hillenbrand et al. 2008)
indicate that the main effects of random errors are on the widths of
the Gaussian core in the $\Delta$\logl\ distributions, and the main
effects of binaries are on the shape of the high side of the
$\Delta$\logl\ distribution.  True age spread, if present, may be
detectable as additional spread in $\Delta$\logl\ present on both the
high and low luminosity sides.  We have found from extensive K-S
testing that when observational errors are modest ($\sim$10\%) and
binarity properties of the underlying population are well understood,
age spreads larger than $\sim$15-20\% can be distinguished from no age
spread or a ``burst" star formation scenario.  We continue to test the
various parameter spaces, including variation of the star formation
history (e.g.  burst, constant, gaussian, linearly/exponentially
increasing or decreasing).  We also find that the fitted slope in
\logl\ vs \logt\ can inform the choice of evolutionary tracks, modulo
the binarity properties.

\section{Findings and Implications}

Pre-main sequence evolutionary tracks:
(1) vary significantly and systematically between theory groups;
(2) under-predict stellar masses by 30-50\%;
(3) under-predict likely low-mass stellar ages by 30-100\%; and 
(4) over-predict likely high-mass stellar ages by 20-100\%.
The above imply large and systematic uncertainties in
both mass and age distributions for young low mass populations, and hence: 
star formation histories in molecular clouds,
disk evolutionary time scales, and
angular momentum evolutionary time scales.

The reality of the observed apparent luminosity spreads in recently
star forming regions can be tested via detailed correlation of the
$\Delta$\logl\ distributions with surface gravity indicators, lithium
abundance measurements, and perhaps soon seismology checks in certain
mass regimes.  HR diagram simulations that account for plausible
error, binary and other astrophysical effects are needed in order to
determine the relevant luminosity spread or $\sigma(L) =
\sqrt{\sigma^2_{observed}(L) - \sigma^2_{understood}(L)}$ that might
then be assessed as a real luminosity spread for inference of any
extended star formation history.

At present, there is only marginal or no strong evidence for moderate
age spreads in young clusters.  However, this conclusion does not
preclude the ``popcorn" effect for cluster star formation history, in
which a few stars form first, preceding the main event, and a few
stars lag, forming last -- just like an episode of popcorn production
relative to the interval between popcorn events in a typical household
microwave oven or other popcorn nursery.

\acknowledgment This presentation has included results established by
my collaborators on various projects: Massimo Robberto and the HST
Orion Treasury team plus Aaron Hoffer on the ONC, Adam Kraus on
multiplicity and Taurus scattered light sources, Catherine Slesnick
plus Davy Kirkpatrick on young star surface gravities, Russel White
plus Amber Bauermeister on HR diagram simulations, and Ann Marie Cody
on brown dwarf pulsations.

%
%

\end{document}